\documentclass{ws-procs975x65}

\begin{document}

\title{PROBING DYNAMICAL DARK ENERGY WITH PRESS-SCHECHTER MASS FUNCTIONS}

\author{MORGAN LE~DELLIOU}

\address{CFTC, Lisbon University,\\
Lisbon, Portugal\\
\email{delliou@cii.fc.ul.pt}}

\begin{abstract}
This project proposes to discriminate in the wealth of models for dark
energy using the formation of non-linear dark matter
structures. In particular,it focuses on structures traced by the
mass function of dark matter haloes.
\end{abstract}


\bodymatter

\section{Introduction}

Among the various models proposed for the explanation of the accelerated
expansion in terms of dark energy (DE), distinctions can be made between
static (cosmological constant) or dynamical (e.g. quintessence), coupled
or uncoupled to dark matter (DM), clustering or unclustering or even
unified (Chaplygin Gas) DE. This wealth of models calls for discriminating
schemes. The goal of this work is to propose a unified analysis extending
previous studies (see Le~Delliou 2006\cite{LeD06}, Manera \& Mota
2006\cite{ManeraMota06} and\cite{MaininiBonomento06} references
therein) of DE impact on dark matter haloes on mass functions, for
confrontation with other DE assessments.%
\begin{table}

\tbl{Choice of scalar field's potentials}
{
\begin{tabular}{>{\raggedright}m{2.5cm}l>{\centering}m{3cm}||>{\centering}m{2cm}l}
\hline 
\multicolumn{3}{c||}{Explored models \cite{LeD06,ManeraMota06}}&
\multicolumn{2}{c}{models under scrutiny}\tabularnewline
\hline 
Model&
Potential V&
Origin&
Model&
Potential V\tabularnewline
\hline 
R.P.&
$\frac{\Lambda_{Q}^{4+\alpha}}{Q^{\alpha}}$&
Historical potential; global SUSY&
Albrecht \& Skordis 2000\cite{AlbrechtSkordis00}&
$\begin{array}{c}
\Lambda_{Q}^{4}\left(\left(\kappa Q-M_{1}\right)^{2}\right.\\
\left.+M_{2}\right)e^{-\lambda\kappa Q}\end{array}$\tabularnewline
SUGRA&
$\frac{\Lambda_{Q}^{4+\alpha}}{Q^{\alpha}}e^{\kappa^{2}\frac{Q^{2}}{2}}$&
SUSY+extra dim.=SUGRA superpot.&
Sahni \& Wang 2000 \cite{SahniWang00}&
$\begin{array}{c}
\Lambda_{Q}^{4}\left(\cosh\left(\lambda\kappa Q\right)\right.\\
\left.-1\right)^{\alpha}\end{array}$\tabularnewline
Ferreira \& Joyce 1998&
$\Lambda_{Q}^{4}e^{-\lambda\kappa Q}$&
extra dim. compactification&
Dodelson \emph{et al.} 2000 \cite{Dodelsonetal00}&
$\begin{array}{c}
\Lambda_{Q}^{4}e^{-\lambda\kappa Q}\left(1+{}\right.\\
\left.\alpha\sin\left(\nu\kappa Q\right)\right)\end{array}$\tabularnewline
Steinhardt \emph{et al.} 1999&
$\Lambda_{Q}^{4}e^{\frac{1}{\kappa Q}}$&
=$\sum$R.P.&
R.P.$\times$F.J.&
$\frac{\Lambda_{Q}^{4+\alpha}}{Q^{\alpha}}e^{-\lambda\kappa Q}$\tabularnewline
\hline
Barreiro \emph{et al.} 2000&
$\begin{array}{c}
\Lambda_{Q}^{4}\left(e^{-\alpha\kappa Q}\right.\\
\left.+e^{-\beta\kappa Q}\right)\end{array}$&
double exponential&
Bertolami \emph{et al.} 2004 \cite{Bertolamietal04}&
$\begin{array}{c}
V_{0}e^{3(\alpha-1)\phi}\times\\
\left(\left(\cosh\left(\frac{\kappa\phi}{2/m}\right)\right)^{\frac{2}{\alpha+1}}\right.\\
\left.+\left(\cosh\left(\frac{\kappa\phi}{2/m}\right)\right)^{\frac{-2\alpha}{\alpha+1}}\right)\end{array}$ \tabularnewline
\hline
\end{tabular}}\label{cap:Choice-of-scalar}
%
%
%\caption{Choice of scalar field's potentials}
%
\end{table}

\vspace{-1cm}

\section{Models and Mass functions}

We model a cosmic fluid with baryons, radiation and, either uncoupled
(and coupled) DM with %
\begin{figure}

\caption{\label{cap:Cumulative-mass-functions}Cumulative mass functions for
different models}

~

~\hspace{-4.3cm}\begin{tabular}{>{\centering}p{10.2cm}c}
\begin{tabular}{c}
~\includegraphics[%
  width=0.79\textwidth,
  height=0.67\textheight,
  keepaspectratio]{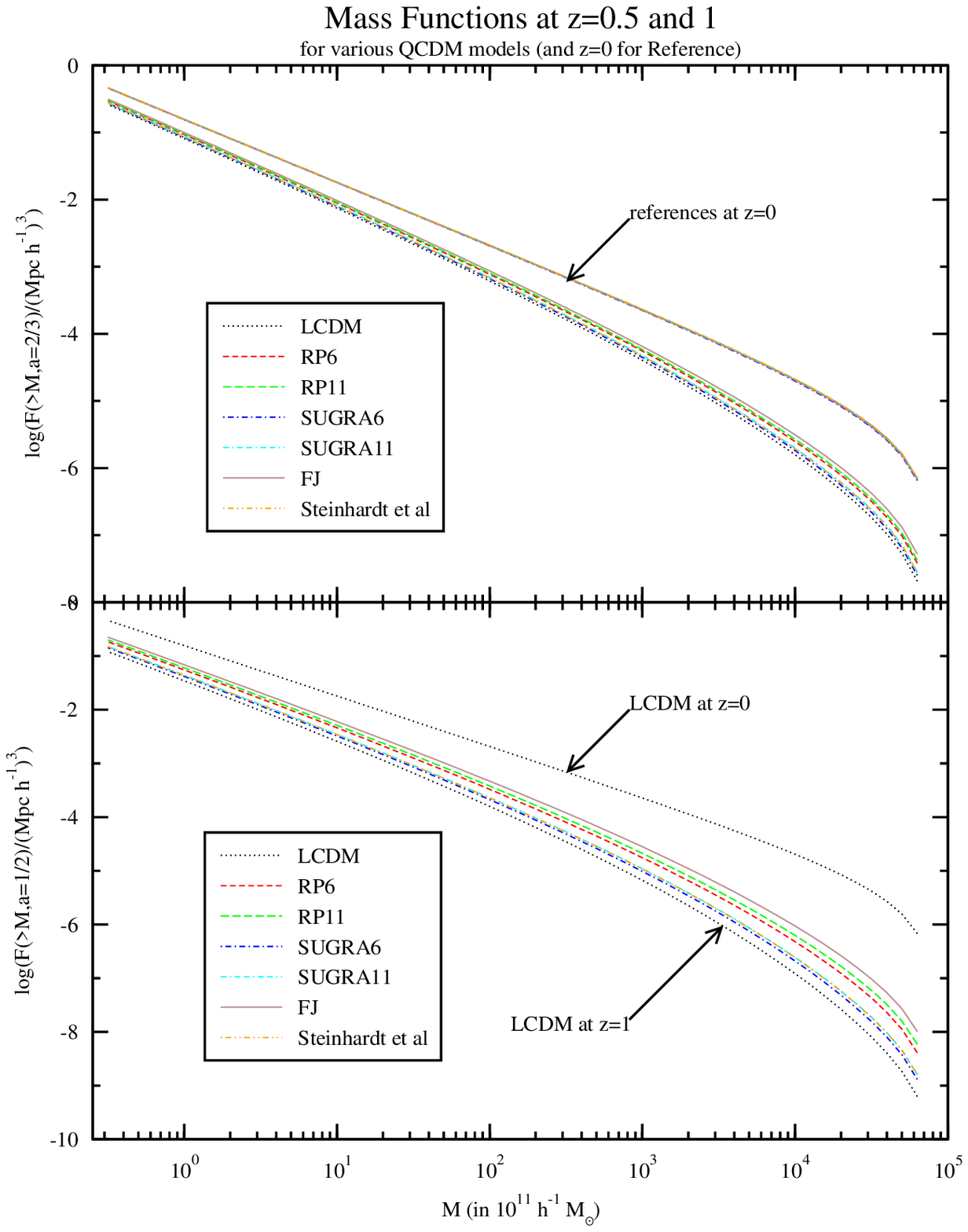}\tabularnewline
\end{tabular}&
\begin{tabular}{>{\raggedright}m{11cm}}
\parbox[c]{11cm}{Barreiro \emph{et al.} model from Manera \& Mota 2006\cite{ManeraMota06}:
\\
A=non Clust., large $\Omega_{cDM_{0}}$\\
B=non Clust., small $\Omega_{cDM_{0}}$\\
C=Clust., large $\Omega_{cDM_{0}}$\\
D=Clust., small $\Omega_{cDM_{0}}$}\tabularnewline
~\vspace{0.9cm}\tabularnewline
~\includegraphics[%
  width=0.72\textwidth,
  keepaspectratio]{Figure1b.eps}\tabularnewline
\end{tabular}\tabularnewline
Le~Delliou 2006\cite{LeD06}&
Manera \& Mota 2006\cite{ManeraMota06}\tabularnewline
\end{tabular}
\end{figure}
a(n) (un)clustering scalar field DE (quintessence) $Q$, (non-)minimally
coupled to DM, or a Chaplygin Gas (GCG) -- DM/DE unified component
-- defined either from $P\propto-\rho^{-\alpha}$ or from a scalar
field mimicking it. We restrict to flat backgrounds, a linear coupling
and model clustering through energy conservation. The models are defined
by their potential (Table \ref{cap:Choice-of-scalar}). We already
studied homogeneous minimal quintessences (left side, upper part{\small \cite{LeD06}})
and a coupled, clustering quintessence (left side, lower part{\small \cite{ManeraMota06}}).
All potentials shown will have clustering and interaction.

We use the top hat spherical collapse to model non-linear structure
formation as a Friedmann sphere with higher, varying curvature. We
extract the linearly extrapolated overdensity as a function of non-linear
collapse scale factor $\delta_{c_{0}}(a_{c})$. This is combined in
a Press-Schechter scheme to get the mass function of large scale structures.
The results obtained so far are presented in Fig.~\ref{cap:Cumulative-mass-functions}.

\section{Conclusions }

Extending previous evidence of DE models impact on DM mass functions,
our results permit the confrontation of several homogeneous models
and the examination of clustering and interacting quintessence. This
have shown that more insights can be drawn from confrontation of several
homogeneous models\cite{LeD06} and that {\tiny }strong effects on
mass function evolution proceed from clustering and interacting quintessence\cite{ManeraMota06}.
Indeed, the spread of $\sim$10\% at $10^{14}\, h^{-1}M_{\odot}$
between mass functions and the hierarchy between models on the lower
($z=1$) panel of the left part of Fig.~\ref{cap:Cumulative-mass-functions}
shows that the method should be most discriminant on clusters scales
and that the impact of $\omega_{Q}$ dominates other effects. Moreover,
its right part entails that, contrary to homogeneous models, DE clustering
increases DM clustering while coupling decreases it. This motivates
our extended study of DE models with mass functions. Some pending
questions remain: our use of Birkhoff's theorem with spherical symmetry
in cosmology may require some mass function corrections; geometric
effects are argued to induce a degeneracy in angular mass functions\cite{SoleviEtal04},
not taking the bias-geometry dependence\cite{Kaiser84} into account.
We are extending our results to other models (Table \ref{cap:Choice-of-scalar};
Chaplygin gas), including clustering and interacting DE. Further developments
are also planned.

\section*{Acknowledgements}

See acknowledgements in \cite{LeD06}. Current work involves J.P.
Mimoso, U. Lisboa, D.F. Mota, U. Heidelberg, C. van de Bruck, U. Sheffield
and O. Bertolami, IST Lisboa.

\end{document}